\documentclass[epj,nopacs]{svjour}
\usepackage{graphicx} 
\usepackage{url}
\usepackage{bm}
\usepackage[normalem]{ulem}
\usepackage{minibox}
\usepackage{color}
\begin{document}

\title{Synchrony breakdown and noise-induced oscillation death in ensembles
of  serially connected spin-torque oscillators}

\author{Michael A. Zaks \inst{1,3} \and Arkady Pikovsky \inst{2,3}}
\institute
{Institute of Physics, Humboldt University of Berlin, Germany,
\email{zaks@physik.hu-berlin.de}
\and 
Institute of Physics and Astronomy, University of Potsdam, Germany,
\email{pikovsky@uni-potsdam.de}
\and
Department of Control Theory, Lobachevsky University Nizhni Novgorod, Russia}

\abstract{
We consider collective dynamics in the ensemble of serially connected 
spin-torque oscillators governed by the Landau-Lifshitz-Gilbert-Slonczewski 
magnetization equation. 
Proximity to homoclinicity hampers synchronization 
of spin-torque oscillators: 
when the synchronous ensemble experiences the homoclinic bifurcation, 
the Floquet multiplier, 
responsible for the temporal evolution of small deviations from the 
ensemble mean, diverges.  Depending on the configuration of the contour,
sufficiently strong common noise, exemplified by stochastic oscillations
of the current through the circuit, may suppress precession of the magnetic
field for all oscillators. We derive the explicit expression for the threshold
amplitude of noise, enabling this suppression.}

\authorrunning{M.A. Zaks, A. Pikovsky}

\titlerunning{Synchrony breakdown and noise-induced oscillation death
in serially coupled STOs}

\date{\today}

\maketitle

\section{Introduction}
\label{sect:intro}

Synchronization transition in systems of coupled oscillators can be considered
as a nonequilibrium order-disorder phase 
transition~\cite{Kuramoto-75,Gupta-Campa-Ruffo-18,Pikovsky-Rosenblum-15}.
Its manifestation is appearance of a macroscopic mean field in the ordered phase,
while in the disordered phase macroscopic mean field vanishes in the thermodynamic
limit or fluctuates at a small level in finite ensembles. This property can be used 
for a coherent summation of the outputs of generators, which being uncoupled have random
phases and thus produce a small output. Recently, this idea has been explored
for spin-torque oscillators (STOs)~\cite{Slavin-Tiberkevich-09,Chen_etal-16}. 
These nanoscale spintronic devices generate microwave oscillations (in the frequency 
range of several GHz), but the output is too weak for applications. 
Thus, one has looked for different schemes of coupling
in order to synchronize the STOs. One possibility is magnetodipolar 
coupling of vortex-based STOs, explored in 
Refs.~\cite{Araujo-Grollier-16,PhysRevB.93.224410,7120163,%
PhysRevB.92.045419,PhysRevB.90.054414,Demidov_etal-14,%
Erokhin-Berkov-14,Lebrun_etal-15,Flovik-Macia-Wahlstrom-16}.
Another popular setup is electric coupling through the common microwave 
current~\cite{Grollier-Cros-Fert-06,georges:232504,Li_Zhou_2010,%
PhysRevB.84.104414,PhysRevB.86.014418,%
Pikovsky_2013,Turtle_etal-17,Tiberkevich_etal-09,Tsunegi_etal-18}.

Studies of serial arrays of STOs have shown that they are not easy to synchronize --
quite often, instead of desired coherent oscillations, 
complex asynchronous or partially synchronous states are 
observed~\cite{sci_rep_2017}. 
To overcome this asynchrony, schemes with additional periodic external 
field~\cite{Subash_etal-15} or with delay in coupling~\cite{Lebrun_etal-17} 
have been suggested. 
One of the goals of this paper is to find out, 
why synchrony is so vulnerable in STO arrays, contrary to predictions
of simple models based on the Kuramoto-type 
equations~\cite{georges:232504,Flovik-Macia-Wahlstrom-16}.
Below, we demonstrate that the reason is in the homoclinic (gluing) bifurcation of the limit 
cycles~\cite{PhysRevB.84.104414,Turtle_etal-13,our_gluing}, 
close to which the transversal instability of the synchronized state becomes enormous. 

In the second part of the paper,
we explore effect of noise on the synchrony. We consider fluctuations in
the common microwave current, which does not directly violate synchrony
and can even facilitate it~\cite{Nakada-Yakata-Kimura-12,Arai-Imamura-17}.
However, presence of a common load makes the effect of noise nontrivial. We demonstrate,
that under certain conditions a strong enough common noise can lead to oscillation death:
a steady state which without noise is unstable, becomes stabilized, so that the oscillations 
of magnetic field disappear.

The layout is as follows: in Sect.~\ref{sect_sto} we briefly explain 
the physical mechanisms and present the governing equations 
for the single spin-torque oscillator. Increase of the current
through this STO destabilizes its state of equilibrium and gives rise
to periodic oscillations. 
In Sect.~\ref{sect_eq_circuits} we introduce three exemplary circuits 
with serially connected identical STOs and discuss 
the onset of oscillations in each circuit.
In Sect.~\ref{sect_oscillations} we show that further evolution leads
through the formation of homoclinic orbits in the partial phase spaces
of the oscillators. The transversal Floquet multiplier,
responsible for the stability of collective oscillatory states, diverges
at the homoclinic bifurcation, resulting in extremely strong instability
of synchronous oscillations.  Finally, in Sect.~\ref{section_noise} we consider
dynamics under the influence of the noisy common current and discuss
the conditions under which the fluctuations of current are able to suppress
the oscillations and effectively restabilize the state of equilibrium.

\section{Spin-torque oscillator: governing equations}
\label{sect_sto}
In a serially connected circuit, interaction of spin-torque oscillators 
takes place by means of the common electric current. 
In this way, instantaneous magnitude of the current becomes
an explicit parameter in dynamical equations of every oscillator. 
Since the current is common, its value is obtained 
through the self-consistent closure of the system. 
Details of the closure depend on the total number of oscillators in the circuit,
on possible heterogeneities in the ensemble and on the circuit configuration: 
presence of capacitors, inductances and other elements. 
However, parameterization in terms of the current can be performed 
for every single unit separately; hence, many relevant dynamical properties 
of the ensemble, including the stability of the equilibrium  states,
can be derived from the equations of motion of the solitary oscillator.

\begin{figure}[h]
\centerline{\includegraphics[width=0.14\textwidth]{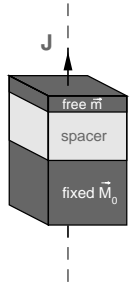}}
\caption{Spin-torque oscillator in the circuit.}
\label{fig_setup}
\end{figure}

Consider a spin-torque oscillator in the circuit. 
In the simplest variant, it comprises two magnetic layers 
separated by the nonmagnetic spacer  (Fig.\ref{fig_setup}).
In the thicker layer the magnetization is constant, 
whereas in the thinner one it can freely rotate. 
When the electric current passes through the thick magnetic layer,  
the electrons interact with magnetic field and become spin-polarized. 
Injected into a thin free magnetic layer, this polarized current 
induces precession of magnetization.
The macroscopic description of this process 
is delivered by the Landau-Lifshitz-Gilbert-Slonczewski magnetization equation. 
Below, we largely follow the notation of~\cite{Li_Zhou_2010}.
For the unit vector  $\vec{m}$ of magnetization 
in the free layer, the LLGS equation reads
\begin{equation}
\frac{d\vec{m}}{dt}=-\gamma \vec{m}\times\vec{H}_{\rm eff}
+\alpha\vec{m}\times\frac{d\vec{m}}{dt}
+\gamma\overline{\beta} J\,\vec{m}\times (\vec{m}\times\vec{M}_0)
\label{llgs}
\end{equation}
where the last term, as shown by Slonczewski~\cite{Slonczewski}, 
characterizes the current-driven spin transfer.
Here,  $\alpha$ denotes the Gilbert damping coefficient, 
$\gamma$ is the gyromagnetic ratio, 
whereas the effective Landau-Lifshitz field $\vec{H}_{\rm eff}$ consists 
of three components: the external magnetic field $\vec{H}_a$, 
the uniaxial anisotropy field $\vec{H}_k$ directed along the axis of
easy magnetization, and the demagnetizing contribution $\vec{H}_{dz}$.
In the spin-transfer term, 
$\vec{M}_0$ is the constant magnetization of the thick fixed layer, 
$\overline{\beta}$ characterizes the material properties of the free layer, 
and finally (but most importantly in our context), 
$J$ is the instantaneous current 
through the element. By means of this current, every oscillator is coupled 
to the circuit and, thereby, to the rest of the ensemble. 
Depending on its design, the circuit can play a role
of the passive load (purely resistive circuit) or, in presence of capacitors
and/or inductances, possess its own degrees of freedom.
Notably,  $J=J(t)$ is, in general, time-dependent:
spin-transfer changes back and forth the magnetoresistance of the spin-torque 
oscillator 
(for quantitative description of this process see~\cite{Grollier-Cros-Fert-06}),
therefore Eq.(\ref{llgs}), taken out of the context of the surrounding 
circuit, is essentially non-autonomous.

On aligning the $x$- and $z$-axes of the coordinate system with directions of, 
respectively, the external field $\vec{H}_a$ 
and the demagnetization field $\vec{H}_{dz}$, 
Eq. (\ref{llgs}) turns into the
coupled equations for the components of $\vec{m}$:
\begin{eqnarray}
\label{cartesian}
\frac{1}{\Gamma}\,\frac{d m_x}{dt}&=& H_{dz} m_y m_z
       +\alpha\left( (H_a+H_k m_x)(m_y^2+m_z^2)\right.\nonumber\\ 
              &&\left.+H_{dz}m_x m_z^2)\right)
        -\overline{\beta} J M_0 (m_y^2+m_z^2)\\
\frac{1}{\Gamma}\,  \frac{d m_y}{dt}&=& 
    H_{dz}m_z (\alpha m_y m_z -m_x) -H_k m_x (\alpha m_x m_y +m_z)
    \nonumber\\
   & &
    -H_a (\alpha m_x m_y+m_z)+\overline{\beta} J M_0 (m_x m_y-\alpha m_z) \nonumber\\
\frac{1}{\Gamma}\, \frac{d m_z}{dt}&=& (H_a+H_k m_x)(m_y-\alpha m_x m_z)
         \nonumber\\
	   & & -\alpha H_{dz}(m_x^2+m_y^2) m_z 
	     + \overline{\beta} J M_0 (m_x m_z+\alpha m_y)\nonumber
\end{eqnarray}
where $\Gamma$ abbreviates the factor $\gamma/(1+\alpha^2)$. 
Choice of the coordinate system implies that 
the coefficients $H_a$ and $H_{dz}$ are positive;
the value of $H_k$, without restrictions, can be viewed as positive as well. 

\subsection{States of equilibrium and their stability}

Since the vector $\vec{m}$  is orthogonal to the  rhs of Eq.(\ref{llgs}), 
its length is conserved, while orientation can vary in time. 
Accordingly, the partial phase space of a single spin-torque oscillator 
is  the two-dimensional spherical surface.  
For a set of $N$ such oscillators, the phase space is a direct sum of $N$ 
spheres, augmented by directions which correspond to independent global 
variables of the circuit  (e.g. voltage).
A look at the equations  (\ref{cartesian}) shows that
a magnetic moment $\vec{m}$, if set parallel to the external field $\vec{H}_a$,
stays constant and preserves orientation: if, originally, the off-field 
components $m_y$ and $m_z$ vanish identically, they will not be excited, 
and precession of  $\vec{m}$ would not arise.
For a single oscillator, there are two such states of equilibrium, 
characterized by  $m_x=\pm 1$. 
For the whole ensemble this implies that every element 
which, in the course of evolution, gets {\em exactly} parallel to $\vec{H}_a$,
remains in that equilibrium position forever
and does not contribute to generation of the electromagnetic field. 
Notably, these states of equilibrium exist independently from 
the circuit composition and from the value of the 
current $J$ through the stack of STO units: 
what is influenced by $J$, is their stability~\footnote{Besides the states 
with $m_x=\pm 1$, there may be other stationary directions of $\vec{m}$. 
Their existence, in contrast, depends on the parameters of the problem and 
on the circuit details; in examples known to us, such states are unstable.}.

Consider, first, a single STO.
There are two possibilities for the equilibrium: 
$\vec{m}$ is oriented either along the external field $\vec{H}_a$ 
or in the opposite direction.
In the equations (\ref{cartesian}) the terms containing the current $J$ 
are proportional to linear and quadratic terms in $m_y,\,m_z$. Therefore, 
linearization near the states with  $m_x=\pm 1$ 
(and, hence, the stability of those states)
depends only on the value of the time-independent component of $J$. 
The equilibrium  value of $J$ is dictated by the configuration of the
circuit into which the oscillators are included.
Below, we will list a few exemplary circuit configurations and explicitly
express the respective values of $J$ through the value of the external 
current $I$; however, until the end of the current section 
we use the symbol $J$ for parameterization of dynamics near the equilibria. 
 
To begin with, we characterize the state with $m_x=1$: magnetization along
the external field $\vec{H}_a$. The characteristic equation reads 
\begin{eqnarray}\label{equi_mx1}
\lambda^2&&+2\lambda\left(\alpha (H_a+H_k+\frac{H_{dz}}{2})-\beta J\right)\\
&&+(1+\alpha^2)\left(\beta^2 J^2+(H_a+H_k)(H_a+H_k+H_{dz})\right)=0.\nonumber
\end{eqnarray}
(here $\beta$ denotes the product $\overline{\beta} M_0$,
and the factor $\Gamma$ is absorbed in the time units).

Since the last term of Eq.(\ref{equi_mx1}) is positive, two eigenvalues 
cannot have opposite signs.
If the current $J$ is absent or sufficiently weak, the equilibrium is stable; 
it gets destabilized in the Hopf bifurcation at
\begin{equation}\label{bif_Hopf}
J=J_H=\frac{\alpha}{\beta}\,(H_a+H_k+\frac{H_{dz}}{2}).
\end{equation}

The second equilibrium, with magnetization directed opposite to the 
external field $\vec{H}_a$ ($m_x$=--1),
looks intuitively unstable. This is indeed true, as long as the current 
$J$ is not too large. The corresponding characteristic equation is 
\begin{eqnarray}\label{equi_mx_min1}
\lambda^2&&+2\lambda\left(\beta J+\alpha (H_k-H_a+\frac{H_{dz}}{2})\right)\\
&&+(1+\alpha^2)\left(\beta^2 J^2+(H_k-H_a)(H_k-H_a+H_{dz})\right)=0.\nonumber
\end{eqnarray}
For $\displaystyle J<J_{st}=\frac{1}{\beta}\sqrt{(H_a-H_k)(H_k-H_a+H_{dz})}$, 
the last term is negative, and the steady state is a saddle. 
At $J=J_{st}$ the pitchfork bifurcation stabilizes this equilibrium. 

For generic values of $H_k,\,H_a,\,H_{dz}$ the value $J_{st}$ exceeds $J_H$ 
by the order of $1/\alpha$. Since the Gilbert damping coefficient $\alpha$ 
is typically of the order of $10^{-2}$~\cite{Sinova_2004}, this ensures 
a broad range of values of $J$ in which one equilibrium state is an unstable 
focus whereas another one is a saddle point, {\em regardless of the design 
of the circuit into which the STO elements are serially included}. 
Within this range of $J$, every unit in a set of identical spin-torque elements 
performs oscillations, and later we will show that at least in some part 
of the range those oscillations cannot be synchronized.

In stack of $N$ STO, each element can occupy any
of two possible equilibrium positions, hence there are altogether 
$2^N$ collective states of equilibrium. 
Consider the configuration with $N_+\geq 0$ oscillators having $m_x=1$ 
and the remaining $N_-=N-N_+$  units with $m_x=-1$.
Eigenvalues that characterize growth/decay of small deviations for
the former oscillators obey Eq.(\ref{equi_mx1}); each of them is $N_+$-times
degenerate. Eigenvalues of the oscillators with $m_x=-1$ are described
by Eq.(\ref{equi_mx_min1}); their degree of degeneracy equals $N_-$.
At $J<J_{st}$ this collective equilibrium possesses $N_-$ real positive 
Jacobian eigenvalues, at $J>J_H$ it possesses $2N_+$
complex eigenvalues with positive real parts.
Therefore, collective ``mixed'' states 
with part of the oscillators aligned with the field $\vec{H}_a$ whereas 
the rest is directed strictly opposite to it (i.e. $N_+N_->0$), 
are unstable at all values of the current $J$. 
As for the ``pure'' states of equilibrium, the 
state in which all $m_x$ are aligned with the field
is stable (unstable) below (above) $J_H$; 
the state with all magnetization vectors antiparallel to the field 
is a saddle with $N$ equal positive eigenvalues 
for $J<J_{st}$ and becomes stable beyond $J_{st}$.

For further progress, we need to know how the value of $J$ is related to 
the control parameters of the setup, i.e. to the total current $I$ 
that flows across the circuit: 
that relation differs over different arrangements of the circuits. 
Before discussing various circuits, is is convenient to lower the order 
of dynamical system, using  the conservation of length of the vector $\vec{m}$
and proceeding from $(m_x,m_y,m_z)$ to spherical angles 
$\theta$ and $\varphi$: $m_x=\sin\theta\cos\varphi$, 
$m_y=\sin\theta\sin\varphi$, $m_z=\cos\theta$.
In the set of $N$ spin-torque oscillators each element is characterized 
by its own instantaneous angles $\theta_i$ and $\varphi_i$; for the $i$-th 
unit, Eq.(\ref{cartesian}) becomes 
\begin{eqnarray}
\label{eq_angular}
  \frac{d\theta_i}{dt}&=&(\alpha H_a-\beta J) 
         \cos\theta_i\cos\varphi_i\nonumber\\
	&& -( H_a+\alpha\beta J)\sin\varphi_i+\alpha S_i -T_i\\
     \sin\theta_i\,\frac{d\varphi_i}{dt}&=&
 -(\alpha H_a-\beta J)\sin\varphi_i\nonumber\\
&& -(H_a+\alpha\beta J)\,\cos\varphi_i\cos\theta_i-S_i-\alpha T_i\nonumber,
\end{eqnarray}
where the symbols $S_i$ and $T_i$ denote, respectively, 
the combinations $(H_{dz}+H_k\cos^2\varphi_i)\sin\theta_i\cos\theta_i$ 
\  and \\ $H_k\sin\theta_i\sin\varphi_i\cos\varphi_i$~\cite{Li_Zhou_2010}.
Compared to (\ref{cartesian}), the time units are rescaled by the
factor  $\Gamma$.

The set of $N$ pairs of Eqs (\ref{eq_angular}) is the main building
block for all circuit
configurations; particularities of circuits enter these equations
as soon as $J$ is expressed through the control parameters of the circuit.

\section{Circuits with serially connected STOs: equations of motion.}
\label{sect_eq_circuits}
Interaction within the set of STOs is mediated by the time-dependent 
common current $J(t)$ through the stack; 
variations of $J(t)$ are caused by variable
magnetoresistance of the units. Spin transfer changes the instantaneous 
resistance of the STO: decreases it, when magnetization in the free layer
is aligned with the external field, and enhances it when the magnetization
includes a component, directed opposite to the field. 
As demonstrated in~\cite{Grollier-Cros-Fert-06}, the value of magnetoresistance 
is a harmonic function of the instantaneous angle 
$\phi_{\vec{m}\vec{M}_0}$ between the magnetizations $\vec{m}$ and 
$\vec{M_0}$, adequately represented by $\cos\phi_{\vec{m}\vec{M}_0}$. 
The lowest value of resistance, $r_p$, is achieved in the case 
when both magnetizations are parallel; 
the highest one, $r_{ap}$, corresponds to antiparallel magnetizations. 
In our configuration, $\vec{M}_0$ is directed along the $x$-axis,
hence $\cos\phi_{\vec{m}\vec{M}_0}=m_x$.
Accordingly, the resistance of the STO is
$$r(t)=\frac{r_p+r_{ap}}{2}\, \big(1-\varepsilon\,m_x(t)\big)$$
where $\varepsilon$ denotes the ratio $(r_{ap}-r_p)/(r_{ap}+r_p)$,
so that $0<\varepsilon<1$.

Below, while treating the stacks of $N$ serially connected STOs, we assume
that the units are identical: they share the values of constants 
$\alpha,\,\beta,\,H_a,\,H_k,\,H_{dz},\, \varepsilon$. 
Due to serial connection of the STOs, the $x$-component
of magnetization is effectively averaged over the stack:
\begin{equation}
\label{resistance_ensemble}
r(t)= \frac{r_p+r_{ap}}{2}\, 
\left(1-\varepsilon\langle m_x\rangle\right)
\end{equation}
with 
\begin{equation}
\label{average_m_x}
\langle m_x\rangle =\frac{1}{N} 
\sum_i^N \sin\theta_i\cos\phi_i . 
\end{equation}
In the collective state of equilibrium with individual vectors of
magnetizations of \textit{all} STOs
directed along (or opposite to) the external field, 
$\langle m_x\rangle=1$ (respectively, $\langle m_x\rangle=-1$).
In the mixed equilibrium state, $\langle m_x\rangle=2N_+/N-1$. 

Consider three exemplary circuits with $N$ serially connected STOs 
where the governing parameter is the constant \textit{external} current $I$: 
the purely resistive load, a circuit with capacitor parallel to the stack,
and a circuit with the $LC$ element.
\begin{figure}[h]
\centerline{\includegraphics[width=0.5\textwidth]{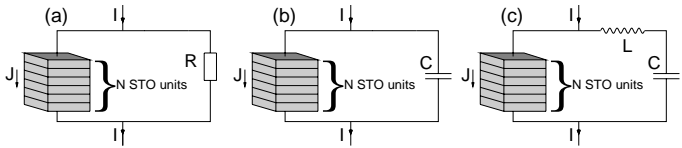}}
\caption{Serially connected STOs in exemplary circuits:\protect\\ 
(a) resistive load, (b) RC circuit, (c) LC circuit.}
\label{fig_circuits}
\end{figure}

\subsection{Resistive load}
In this configuration, sketched in the left panel of Fig.~\ref{fig_circuits}, 
an ohmic load $R$ is set parallel to the STO stack. 
The common current through the STOs is 
\begin{equation}
\label{J_resistive}
J(t)= I\frac{R}{R+r(t)}=\frac{I}{1+\rho(1-\varepsilon\langle m_x\rangle)},
\end{equation}
where  $\langle m_x\rangle$ is given by (\ref{average_m_x})
and $\rho$ is the ratio of resistances:
$$ \rho=\frac{r_p+r_{ap}}{2\,R}.$$ 
On substituting (\ref{J_resistive}) into Eq.(\ref{eq_angular}), 
we obtain a set of 2$N$ equations
\begin{eqnarray}
\label{eq_resistive}
 \frac{d\theta_i}{dt}&=&\left(\alpha H_a
         -\frac{I\,\beta}{1+\rho(1-\varepsilon\langle m_x\rangle)}\right) 
         \cos\theta_i\cos\varphi_i\nonumber\\
	 &&-\left( H_a+\frac{\alpha I\,\beta}
	    {1+\rho(1-\varepsilon\langle m_x\rangle)}\right)\sin\varphi_i
             +\alpha S_i -T_i,\nonumber\\
&&\\
  \sin\theta_i\,\frac{d\varphi_i}{dt}&=&
 -\left(\alpha H_a-\frac{I\,\beta}
              {1+\rho(1-\varepsilon\langle m_x\rangle)}\right)\sin\varphi_i
	     -S_i-\alpha T_i \nonumber\\
 &&-\left(H_a+\frac{\alpha I\,\beta}
              {1+\rho(1-\varepsilon\langle m_x\rangle)}\right)\,
    \cos\varphi_i\cos\theta_i\nonumber\\
& & i=1,\ldots,N.\nonumber
\end{eqnarray}
The resistive circuit is passive: there are no independent variables besides
2$N$ angular coordinates of the STOs.
Substituting the corresponding
values of $\langle m_x\rangle$ into (\ref{J_resistive})
renders equilibrium value of the current $J$:
$$\displaystyle J=\frac{I}{1+\rho(1\pm\varepsilon)}$$  
with the sign before $\varepsilon$ in the denominator being
taken opposite to the sign of $\langle m_x\rangle$.
By inserting this expression
into Eqs (\ref{equi_mx1},\ref{bif_Hopf},\ref{equi_mx_min1}), we relate
the eigenvalues of the equilibria and the threshold $I_H$ 
of the Hopf bifurcation to the external current $I$. 
 
\subsection{Circuit with a capacitor}
Introduction of a capacitor parallel to the stack 
(central panel of Fig.~\ref{fig_circuits})
raises the order of the dynamical system. In this configuration 
the current $J$ through the stack is related to the external current $I$
by $$J(t)\,=\,I-C\,\Gamma\, \frac{dV}{dt}$$
with $C$ denotes the capacitance and $V$ is the voltage 
difference on the stack; the factor $\Gamma$ 
translates the derivative into the 
rescaled time units of Eqs(\ref{eq_angular}).
On combining this with  $V=r(t)\,J(t)$ and introducing the dimensionless 
voltage  $\displaystyle u=\frac{2V}{I\,(r_p+r_{ap})}$, 
equations (\ref{eq_angular}) turn into
\begin{eqnarray}
\label{eq_rc}
  \frac{d\theta_i}{dt}&=&(\alpha H_a-\frac{\beta I u}
      {(1-\varepsilon\langle m_x\rangle)})\cos\theta_i\cos\varphi_i\nonumber\\
	&& -( H_a+\frac{\alpha\beta I u}
	 {(1-\varepsilon\langle m_x\rangle)})\sin\varphi_i+\alpha S_i -T_i
           \nonumber\\
	   &&\\
  \sin\theta_i\,\frac{d\varphi_i}{dt}&=&
 -(\alpha H_a-\frac{\beta I u}
    {(1-\varepsilon\langle m_x\rangle)})\sin\varphi_i-S_i-\alpha T_i\nonumber\\
    &&-(H_a+\frac{\alpha\beta I u}{(1-\varepsilon\langle m_x\rangle)})\,
      \cos\varphi_i\cos\theta_i,\nonumber\\
& & i=1,\ldots,N.\nonumber
\end{eqnarray}
with additional dynamical relation
\begin{eqnarray}
\label{eq_u}
\frac{du}{dt}&=&\omega 
\left(1-\frac{u}{(1-\varepsilon\langle m_x\rangle)}\right)  
\end{eqnarray}
where $\omega$ denotes the parameter combination (inverse characteristic time)
\begin{equation}
\label{def_omega}
\omega=\frac{2(1+\alpha^2)}{\gamma\,(r_{p}+r_{ap})\,C}.
\end{equation}
Altogether, dynamics is governed by $2N+1$ equations.
 
Since in this circuit the segment parallel to the stack bears 
no ohmic resistance, for every steady state the current $J$ through 
the stack is the whole external current $I$.
Therefore, while determining stability and eigenvalues of the 
equilibria, the symbol $J$ 
in the Eqs (\ref{equi_mx1},\ref{bif_Hopf},\ref{equi_mx_min1}), 
should be substituted by $I$.

\subsection{LC-circuit}
Including the inductance $L$ and capacitance $C$ parallel to the stack of STOs
(right panel of Fig.~\ref{fig_circuits}) turns the circuit equation into
\begin{equation}
LC\Gamma^2\,\frac{d^2V}{dt^2} +V =r(t)\,J(t)=r(t)
\left(I-C\,\Gamma\frac{dV}{dt}\right).
\label{eq_circuit}
\end{equation}

On combining (\ref{eq_angular}) with (\ref{eq_circuit}),
we arrive at the system 
of $(2N+2)$ ODEs~\cite{Pikovsky_2013}:
\begin{eqnarray}
\frac{d\theta_i}{dt}&=
    &\cos\theta_i\cos\varphi_i\left(\alpha H_a-\beta I(1-w)\right)\nonumber\\  
    &&-\sin\varphi_i\left( H_a+\alpha\beta I(1-w)\right)
	   +\alpha S_i -T_i\nonumber\\
\label{eq_sto}
\sin\theta_i\,\frac{d\varphi_i}{dt}&=&
         -\sin\varphi_i\left(\alpha H_a-\beta I(1-w)\right) -S_i-\alpha T_i\nonumber\\
	 &&-\cos\varphi_i\cos\theta_i\left(H_a+\alpha\beta I(1-w)\right)
	\nonumber\\
&&\\[-3ex]	
\frac{du}{dt}&=&\omega\, w\nonumber\\
\frac{dw}{dt}&=&\frac{\Omega^2}{\omega}\, 
   \bigg((1-w)\,(1-\varepsilon\langle m_x\rangle)\,
   -u\bigg)\nonumber
\end{eqnarray}
where the variable $u$, like above in Eq.(\ref{eq_u}), 
is the rescaled voltage $V$, 
the variable  $w=\displaystyle\frac{C\,\Gamma}{I}\frac{dV}{dt}$ is 
the rescaled time derivative of $V$,
the parameter $\omega$ is defined in (\ref{def_omega}),
and the additional characteristics of the circuit is its
eigenfrequency $\Omega$ (expressed in units of rescaled time): 
$$\Omega=\frac{1}{\Gamma\,\sqrt{L\,C}}.$$
For this configuration, like in the previous case, $J(t)=I-C\,\Gamma\,dV/dt$, 
therefore $J$ in the characteristic equations 
(\ref{equi_mx1},\ref{equi_mx_min1}) should be directly substituted by
external current $I$; in particular, the threshold of the 
Hopf bifurcation $I_H$ equals to (\ref{bif_Hopf}).


Further configurations of the circuit can be treated along the same lines:
combination of Kirchhof equations that describe the circuit dynamics
with a set (\ref{eq_angular}) of $2N$ equations  for individual
oscillators.\\[3ex]

\section{From the Andronov-Hopf bifurcation to homoclinics and beyond.}
\label{sect_oscillations}
\begin{figure}[h]
\centerline{\includegraphics[width=0.5\textwidth]{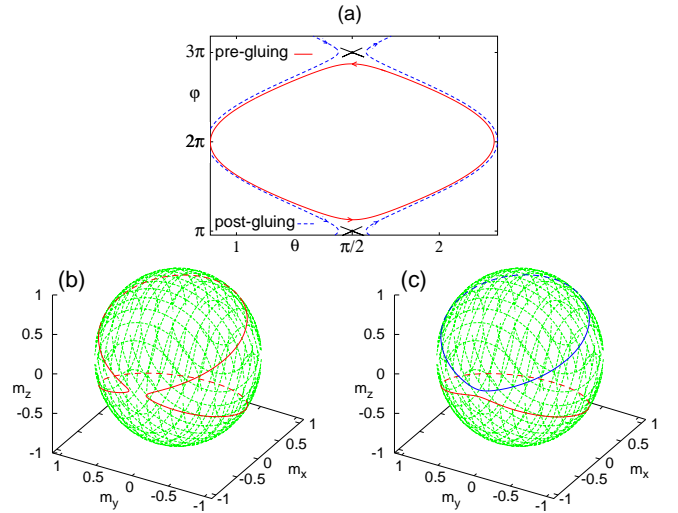}}
\caption{Recombination of attracting orbits in the gluing bifurcation.
Shape of attracting trajectories in the plane (a) 
and spherical (b,c) projections.
Dashed curves: orbit segments on the reverse side of the sphere.
(b) unique limit cycle before the gluing; 
(c) two symmetric limit cycles after the gluing.}
\label{fig_gluing}
\end{figure}

In absence of the external current $I$ there is no precession of magnetic 
field: each STO is at the stable equilibrium with $m_x=+1$. 
For a single unit, increase of $I$ across the bifurcational value (\ref{bif_Hopf})
leads to the onset of periodic oscillations. 
In all three exemplary setups the same bifurcation scenario takes place:
when the current $I$ is increased, the oscillation grows in amplitude
and undergoes the homoclinic bifurcation at which it becomes bi-asymptotic to
the saddle equilibrium with $m_x=-1$. Due to the symmetry of the  governing
equation (\ref{cartesian}) with respect to simultaneous change of sign of
the off-field components $m_y$ and $m_z$, homoclinic orbits exist in pairs, 
therefore a periodic solution does not disappear at homoclinics; instead, 
periodic states recombine in the course of the so-called 
``gluing bifurcation''~\cite{gluing}\footnote{A technical condition that
guarantees stability of recombining closed orbits
is the negative sum of two leading eigenvalues of
Jacobian at the saddle equilibrium; this holds in the case of
the considered STOs~\cite{our_gluing}.}. Transformation of the attracting 
trajectory in that bifurcation is sketched in Fig.~\ref{fig_gluing}.

\begin{figure}[h]
\centerline{\includegraphics[width=0.5\textwidth]{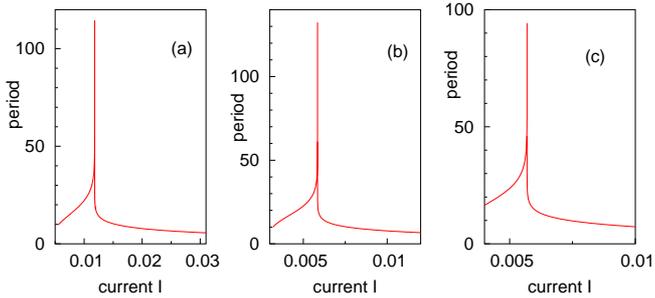}}
\caption{Period of oscillations for an individual STO: 
(a) resistive load, (b) RC circuit, (c) LC circuit.
Common parameter values: $H_{dz}=1.6, H_k=0.05, H_a=0.2$,
$\alpha$=0.01, $\beta$=10/3, $\varepsilon$=0.3. 
Resistive: $\rho$=1; RC and LC: $\omega$=1; LC: $\Omega=1.5$.}
\label{fig_periods}
\end{figure}

Divergence of period $T$ at the bifurcational parameter value $I_{\rm hom}$,
shown in Fig.~\ref{fig_periods},
follows the logarithmic law: $T~\sim -\log|I-I_{\rm hom}|$. The prefactors
on different sides from $I_{\rm hom}$ differ due to the change of the orbit 
shape at the bifurcation~\cite{our_gluing}:
As seen in the panels (b) and (c) of Fig.~\ref{fig_gluing}, 
the cycle of oscillation below $I_{\rm hom}$ is unique and includes two
passages near the saddle point. In contrast, beyond the critical value there are two
symmetric orbits, each of them containing 
just one passage near the saddle per rotation period.

For an isolated STO, the periodic orbit is stable throughout the whole 
parameter range of its existence. For an ensemble of STOs, the synchronous
periodic state in which all units share the instantaneous values of $\theta$
and $\varphi$, is obviously a solution as well; however, it may be unstable with
respect to perturbations that disturb the coincidence of coordinates 
in the oscillating cluster.
In this situation, stability of the periodic state can be characterized 
in terms of the so-called ``evaporation multiplier''
$\mu_{\rm e}$  that characterizes stability of the synchronous cluster 
against ``evaporation''  of its constituents, by quantifying within a period 
of oscillations the growth factor of the distance between the cluster 
and an infinitesimally displaced unit~\cite{evaporation}.
The value of $\mu_{\rm e}$ is recovered from the solution
on the time interval $(0,T)$
of the equation in normal variations near the synchronous trajectory; in those 
linearized equations, the contribution of perturbed unit into the global field
is neglected. Within this setup, $\mu_{\rm e}$ is the leading multiplier 
of the monodromy matrix (in our case, with each unit having two coordinates 
$\theta_i$ and $\varphi_i$, this is a $2\times 2$ matrix).
If  $|\mu_{\rm e}|<1$, the cluster is stable with respect to 
splitting~\footnote{This inequality guarantees return to the cluster of 
sufficiently weakly displaced units but does not imply global 
attractivity of the cluster: 
according to numerics, even when the inequality is fulfilled,
setting the STOs at random initial conditions almost never ends up 
with convergence of all oscillators 
to the synchronous limit cycle~\cite{Pikovsky_2013}.}.
  
\begin{figure}[h]
\centerline{\includegraphics[width=0.49\textwidth]{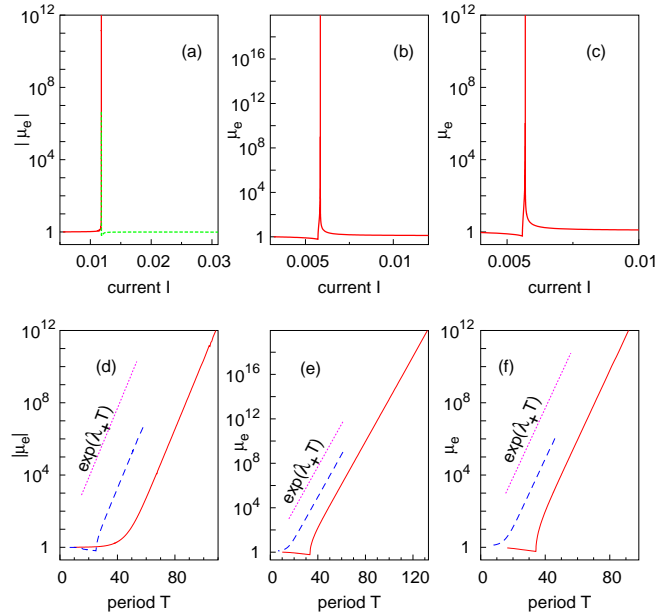}}
\caption{Evaporation multiplier for the synchronous periodic oscillation. 
Left column (a,d): resistive load, middle column (b,e): RC circuit, 
right column (c,f): LC circuit.
Parameter values: see Fig.~\ref{fig_periods}. 
In the panel (a) the dashed part of the curve (to the right from the peak)
corresponds to negative values of $\mu_{\rm e}$.
In the bottom row, solid and dashed curves correspond, respectively, 
to the current ranges below $I_{\rm hom}$ and above $I_{\rm hom}$;
dotted straight lines show  plots of $\exp(\lambda_+ T)$.}
\label{fig_multipliers}
\end{figure}

Numerically recovered dependence $\mu_{\rm e}(I)$ for exemplary configurations
is plotted in Fig.~\ref{fig_multipliers}.  The common feature for all circuits 
is apparent strong divergence of $\mu_{\rm e}(I)$ at $I=I_{\rm hom}$.
Otherwise, stability differs for different setups. For the resistive load 
(curve in Fig.\ref{fig_multipliers}a) the oscillating cluster becomes 
unstable with respect to splitting 
immediately after its birth in the Andronov-Hopf bifurcation
(recall that for a solitary unit the periodic orbit stays asymptotically stable);
divergence at $I=I_{\rm hom}$ is followed by the parameter interval
in which $\mu_{\rm e}$ is negative, and, somewhat later, 
by stabilization of synchrony (the multiplier enters the unit circle 
at  $\mu_{\rm e}=-1$).
The clusters in RC and LC contours, in contrast, are stable near the birth of 
the periodic solution, lose stability shortly before the homoclinic bifurcation
and remain (weakly) unstable for all values of $I$ 
beyond the bifurcation. The most remarkable effect is 
the extremely sharp growth of $\mu_{\rm e}(I)$ 
for periodic solutions close to homoclinicity;
there, the distance in the phase space between the cluster
and the detached unit grows by many orders of magnitude within a single turn
in the phase space.
This effectively prohibits existence of stable synchronous oscillations in the
adjacent parameter range.

The singularity of the evaporation multiplier at homoclinicity 
is enrooted in the divergence of period.
Consider linearization of the flow near the synchronous time-dependent 
trajectory. To compute $\mu_{\rm e}$, the initial disturbance $x(0)$ 
should be set on the appropriate eigenvector  of the monodromy matrix;  
then $\log\mu_{\rm e}$=$\log \big(\|x(T)\|/\|x(0)\|\big)$
$\approx \int_0^T \lambda(t) dt$ 
where $\lambda(t)$ is the leading eigenvalue of the instantaneous
Jacobian matrix. Near homoclinicity, the system spends the prevalent 
proportion of time in the very slow motion
across the vicinity of the saddle point where $\lambda(t)$ is virtually 
indistinguishable from the positive eigenvalue $\lambda_+$ at the saddle: 
the larger root of Eq.(\ref{equi_mx_min1}). 
Therefore the integral
(and with it, the evaporation multiplier) is dominated by $\exp(\lambda_+ T)$.
In Fig.~\ref{fig_multipliers} where the evaporation multiplier 
is plotted versus the period of the orbit, 
the dependence $\mu_{\rm e}\sim\exp(\lambda_+ T)$
is doubtless. Notably, the prefactor before the exponent at the pre-homoclinic 
branch is the squared prefactor at the post-homoclinic branch (cf. its double 
distance from the dotted line in the logarithmic vertical scale 
of the bottom plots); 
this owes to the fact that the periodic orbit traverses the region,
\textit{non-adjacent} to the saddle, twice below $I_{\rm hom}$ 
but once above $I_{\rm hom}$.

The exponential growth of the evaporation multiplier near homoclinicity
is generic: at the critical parameter value, 
local dynamics near the synchronous limit cycle occurs 
in the subspace that is tangential to the plane in which the equilibrium, 
participating in the homoclinic bifurcation, has its local unstable manifold. 
During the long epoch in which the motion is directed along that manifold, 
generic distances grow as $\exp({\lambda_+ t})$.
The same arguments should ensure the instability of synchronous 
nearly homoclinic one-cluster oscillation in every other setup 
with generic global coupling of units.\footnote{The effect 
can be reversed with the help of the specially tailored non-generic scheme of 
coupling to the global field: if the coupling involves only the coordinates 
corresponding to the local stable manifold of the saddle, the relevant
eigenvalue becomes negative. Accordingly, the evaporation 
multiplier exponentially shrinks as a function of the growing period $T$, and 
at the bifurcation parameter value the synchronous cluster becomes superstable!
In the current setup of serially coupled STOs, however, 
this seems hardly feasible.}

In the situations that lack the symmetries ensuring the gluing of periodic 
orbits at the saddle point, the ``usual'' homoclinic bifurcation takes place, 
with periodic state existing only on one side of the bifurcation parameter 
value.  In accordance with the above reasoning, this collective periodic 
solution should lose stability with respect 
to splitting of the synchronous cluster well before the homoclinicity.

Remarkably, destabilization of synchronous states close to homoclinic 
trajectories has a counterpart in the dynamics of distributed systems.
In many translationally invariant spatial systems governed by partial
differential equations, evolution of spatially homogeneous solutions is
finite-dimensional. Certain types of attractors for such finite-dimensional
dynamics, including the homoclinic trajectories and temporally periodic 
solutions close to homoclinic orbits, were shown to be generically unstable 
with respect to spatial perturbations in the form of longwave 
modulation~\cite{Coullet_Riesler_2000,Riesler_2001}. The
situation discussed in the present manuscript is reminiscent of that effect: 
here, the ensemble of finite size replaces the continuum that is present
in the PDE context.
In both cases, the uniform (synchronous) dynamics is described by a
low-dimensional set of equations that in appropriate parameter ranges possess
attracting periodic solutions close to the homoclinic trajectories, and in both
cases these regimes yield to perturbations that disturb the uniformity.  

\begin{figure}[h]
\centerline{\includegraphics[width=0.5\textwidth]{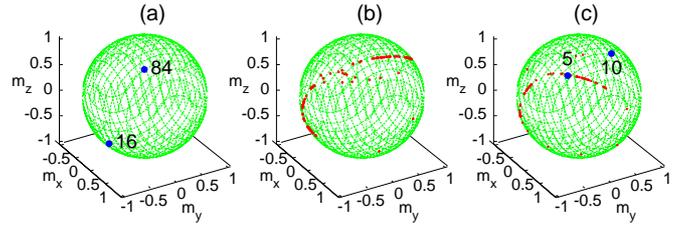}}
\caption{Snapshots of collective states in the LC circuit with $N$=100
 STOs.
Filled red circles: positions of individual oscillators. Filled blue circles:
clusters of oscillators (numbers indicate respective populations.)
(a) periodic state with two clusters; (b) disperse state where all STOs
are disjoint; (c) chimera-like state, combining clusters with isolated
oscillators.}
\label{fig_deterministic}
\end{figure}

Numerical experiments with the STO ensembles 
beyond the threshold values of constant current $I$
in different circuit configurations disclose, mostly, 
complicated dynamical states with various 
degrees of asynchrony~\cite{Pikovsky_2013,sci_rep_2017};
in Fig.~\ref{fig_deterministic} we exemplify a few of them 
by projecting all magnetization vectors onto the same spherical surface.
Within this representation, in the course of temporal evolution
the instantaneous states of the units  typically
move along narrow ring-shape bands.

\section{Action of common noise}
\label{section_noise}
From the point of view of applications, a reasonable way to interfere
into dynamics is to introduce temporal variations for the external
current $I$. Since the same $I(t)$ is perceived by all STO units, it
can be viewed as a common time-dependent signal which affects
the whole  ensemble.
Below we restrict ourselves to the case of modulation with white Gaussian
noise: $I(t)= I_0\left(1+\sqrt{2 D}\,\xi(t)\right)$; 
in this parameterization,  $I_0$ renders the time-average value of the current, whereas 
$D$ is the intensity of the $\delta$-correlated Gaussian random variable
$\xi(t)$.

\subsection{Governing equations}

\subsubsection{Resistive case}
We begin with the stack of STO with resistive load.

On substituting the expression for the modulated current into 
Eq.(\ref{eq_resistive}), we arrive at the set
\begin{eqnarray}
\label{eq_resistive_noisy}
 \frac{d\theta_i}{dt}&=&\left(\alpha H_a
       -\frac{\beta I_0(1+\sqrt{2D}\,\xi(t))}
              {1+\rho(1-\varepsilon\langle m_x\rangle)}\right) 
        \cos\theta_i\cos\varphi_i\nonumber\\
	&& -\left( H_a+\frac{\alpha\beta I_0(1+\sqrt{2D}\,\xi(t))}
	    {1+\rho(1-\varepsilon\langle m_x\rangle)}\right)\sin\varphi_i
             +\alpha S_i -T_i,\nonumber\\
&&\\
  \sin\theta_i\,\frac{d\varphi_i}{dt}&=&
 -\left(\alpha H_a-\frac{\beta I_0(1+\sqrt{2D}\,\xi(t))}
      {1+\rho(1-\varepsilon\langle m_x\rangle)}\right)\sin\varphi_i
      -S_i-\alpha T_i
      \nonumber\\
 &&-\left(H_a+\frac{\alpha\beta I_0(1+\sqrt{2D}\,\xi(t))}
              {1+\rho(1-\varepsilon\langle m_x\rangle)}\right)\,
    \cos\varphi_i\cos\theta_i,\nonumber\\
& & i=1,\ldots,N.\nonumber
\end{eqnarray}
The random variable $\xi(t)$ enters these equations at 4$N$ places,
assuming the \textit{same} value in all of them.

\subsubsection{STO stack with CR circuit}

In the circuit with the capacitor, introduction of the modulation 
of the current does not change the $2N$ governing equations (\ref{eq_rc}) 
for $N$ individual spin-torque oscillators (up to replacing the symbol $I$
by constant $I_0$).
The only modification concerns the equation for the voltage (\ref{eq_u})  
which now reads 
\begin{eqnarray}
\label{eq_u_noise}
\frac{du}{dt}&=&\omega 
\left(1+\sqrt{2D}\,\xi(t)-\frac{u}{(1 -\varepsilon\langle m_x\rangle)}\right)  
\end{eqnarray}
with $D$, like above, being the intensity of the Gaussian white noise $\xi(t)$.
Accordingly, the common noise directly influences dynamics only via
the global variable $u$.

\subsubsection{STO stack with LC circuit}

For time-dependent current $I\left(1+\sqrt{2D}\xi(t)\right)$, 
the ensemble is governed by equations
\begin{eqnarray}
\label{LC_eq_stoch}
\frac{d\theta_i}{dt}&=&U(t)\,\cos\theta_i\cos\varphi_i 
         - W(t)\,\sin\varphi_i
           +\alpha S_i -T_i\nonumber\\
\label{eq_sto_1}
\sin\theta_i\,\frac{d\varphi_i}{dt}&=&
     -U(t)\,\sin\varphi_i  - W(t)\,\cos\varphi_i\cos\theta_i -S_i-\alpha T_i
     \nonumber\\
\frac{du}{dt}&=&\omega\, w\\
\frac{dw}{dt}&=&\frac{\Omega^2}{\omega}\, 
  \bigg(
   (1+\sqrt{2D}\xi(t)-w)\nonumber\\
   &&\hspace*{0.7cm}(1-\frac{\varepsilon}{ N}\sum_j\sin\theta_j\cos\varphi_j)
     -u\bigg)\nonumber
\end{eqnarray}
with
explicit functions of time\\[-3.5ex]
\begin{eqnarray}
U(t)&=&\alpha H_a-\beta(1-w) I_0 \left(1+\sqrt{2D}\xi(t)\right),\nonumber\\
W(t)&=&H_a+\alpha(1-w)\beta I_0 \left(1+\sqrt{2D}\xi(t)\right).\nonumber
\end{eqnarray}
Thereby, in the system of equations the same value of the random variable $\xi$
is employed at $4N+1$ places.

\subsection{Collective dynamics in presence of common noise: phenomenology.}
\label{subsec:phenomenology}
Ensemble dynamics, recovered at $D>0$ by
numerical integration of the stochastic equations of motion 
for $N$=100 and $N$=200 STOs at the values 
of average current beyond the
threshold of the Hopf bifurcation, reminds, in most of the cases, 
disperse states in the deterministic setup. Neither durable states with 
synchronous oscillations of all or, at least, of  the bulk of the elements,  
nor persistent clusters were observed.
In all studied types of circuits, simulations at low values of $D$ 
feature asynchronous dynamics of individual STOs;
instantaneous states of magnetization vectors form bands on the surface
of the unit sphere; when the noise intensity is raised, the bands become wider, 
and now and then separate units temporarily leave the bands, performing large 
excursions over the sphere. 
This behavior seems to be largely insensitive to the initial conditions: 
the same kind of dynamical states evolves from the narrow distributions
near particular points of the sphere and from random homogeneous scattering 
over the spherical surface. 

\subsubsection{Serial stacks of STO with resistive and RC load: 
intermittent alignment with the
external field at high intensity of common noise.}
\begin{figure}[h]
\centerline{\includegraphics[width=0.4\textwidth]{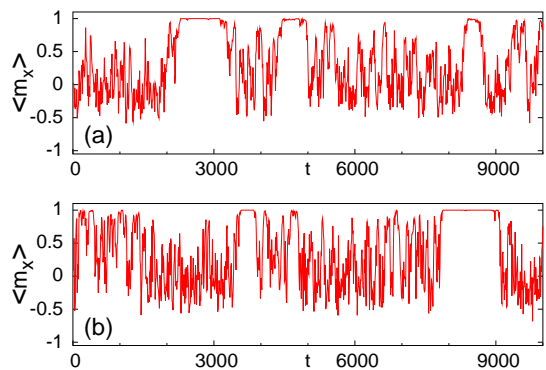}}
\caption{Intermittent alignment to the external magnetic field
at high intensities of common noise.
$I_0=0.01, D=40$.\protect\\
(a) Circuit with purely resistive load, $\rho$=1.
(b) Circuit with RC load, $\omega$=1.
Other parameter values: see Fig.~\ref{fig_periods}. }
\label{fig_intermittency}
\end{figure}

At very large amplitudes of noise, $D\gg 1$, temporal evolution in the stacks
with purely resistive load and with the RC load 
displays a certain kind of intermittency.
From time to time, all magnetization vectors align themselves to the
permanent external field; on the surface of the sphere this is seen as 
temporary contraction of the ensemble to the equilibrium point with $m_x$=1. 
In the plots of $m_x(t)$ (Fig.~\ref{fig_intermittency}), these stages 
are represented by horizontal plateaus.

Recall that  in this range of values of the average current, 
the state of magnetization along the external field is unstable. 
The plateaus at $m_x$=1 owe to repetitive macroscopic segments 
of time in which the local running average over $\xi(t)$ is sufficiently negative, 
so that the real parts of the instantaneous leading Jacobian eigenvalues 
at the steady state are temporarily driven deep into the negative domain. 
This ensures short-term sustainment of the unstable equilibrium.

\subsubsection{Serial stacks of STO with LC load: 
ensemble contracts to a point.}
\begin{figure}[h]
\centerline{\includegraphics[width=0.5\textwidth]{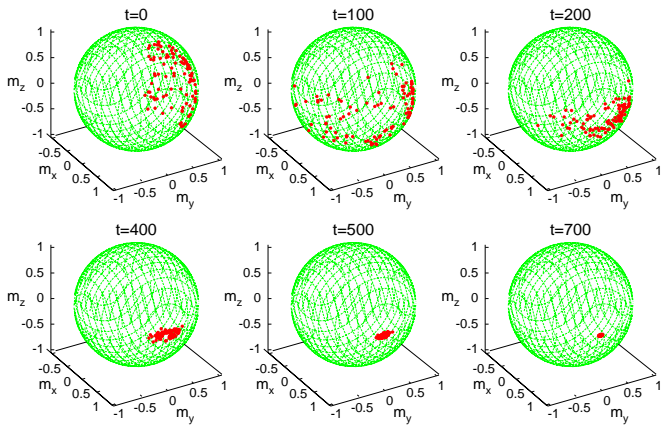}}
\caption{Snapshots of collective states in the LC circuit with $N$=100
 STOs at different time values.
Filled red circles: positions of individual oscillators.}
\label{fig_decay}
\end{figure}

In the case of the STO stack included into the LC circuit, alignment with
the external field becomes permanent. 
In the range of moderate noise values, 
magnetic moments of all STOs, 
regardless of the ensemble size and of their initial orientation, 
gradually converge to the state with $m_x$=1: all of them  become parallel 
to the  external magnetic field.
For the ensemble of 100 STO  units, subsequent stages of the evolution 
on the unit sphere are shown in Fig.~\ref{fig_decay}.

\begin{figure}[h]
\centerline{\includegraphics[width=0.5\textwidth]{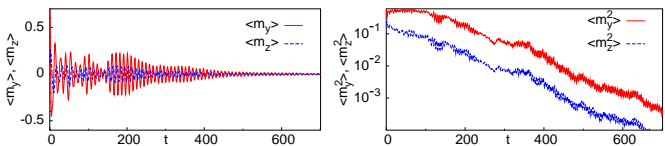}}
\caption{Temporal evolution of the mean off-field components 
at $I=0.01, D=0.5$ with $N$=100.
Other parameter values: see Fig.~\ref{fig_periods}(c).}
\label{fig_off_field}
\end{figure}

Instantaneous individual magnetizations, initially randomly scattered
over large areas of the sphere, gradually evolve into the broad fuzzy 
(and non-uniformly populated) ring-shaped band revolving around 
the equilibrium configuration $m_x=1$; 
as time goes on, the band contracts and finally shrinks to a point.
Convergence to  $m_x$=1 implies gradual vanishing of the 
off-field components $m_y$ and $m_z$. To visualize this process,
we plot in the left panel of Fig.~\ref{fig_off_field} 
temporal dependencies for the averaged
values $\langle m_y\rangle$ and $\langle m_z\rangle$.
In the course of time, irregular evolution of off-field averages
is replaced by ordered oscillatory decay. Since 
smallness of $\langle m_y\rangle$ and $\langle m_z\rangle$ does not exclude
ring-like configurations of units, 
we present in the right panel of Fig.~\ref{fig_off_field}
the evolution of mean squared characteristics 
$\langle m_y^2\rangle$ and $\langle m_z^2\rangle$.

In the deterministic setup 
this range of the current $I$
corresponds  to the unstable collective equilibrium with $m_x$=1 
and to angular precession of the magnetic moment.
We see that in the LC configuration of the circuit the action of sufficiently 
strong common noise is able to stabilize the equilibrium 
and to suppress precession completely. \
Notably, this phenomenon bears the threshold character: 
for it to occur, the intensity of noise should exceed the certain level. 

\subsection{Collective dynamics in presence of common noise: local analysis.}

Since the same noise acts upon all identical units, the system of stochastic 
equations admits a synchronous solution in which the instantaneous values 
of all $\theta_i$ as well as of all $\varphi_i$ coincide.
Stability of this solution is characterized in terms
of the transversal Lyapunov exponent: the average
growth rate of disturbances, splitting the synchronous dynamics.
Below we study dependence of this characteristics on the average current 
$I_0$  and noise intensity $G$ for all considered types of STO circuits.

\subsubsection{STO with resistive and RC load: absence of 
noise-induced large-scale stabilization}

\begin{figure}[h]
\centerline{\includegraphics[width=0.49\textwidth]{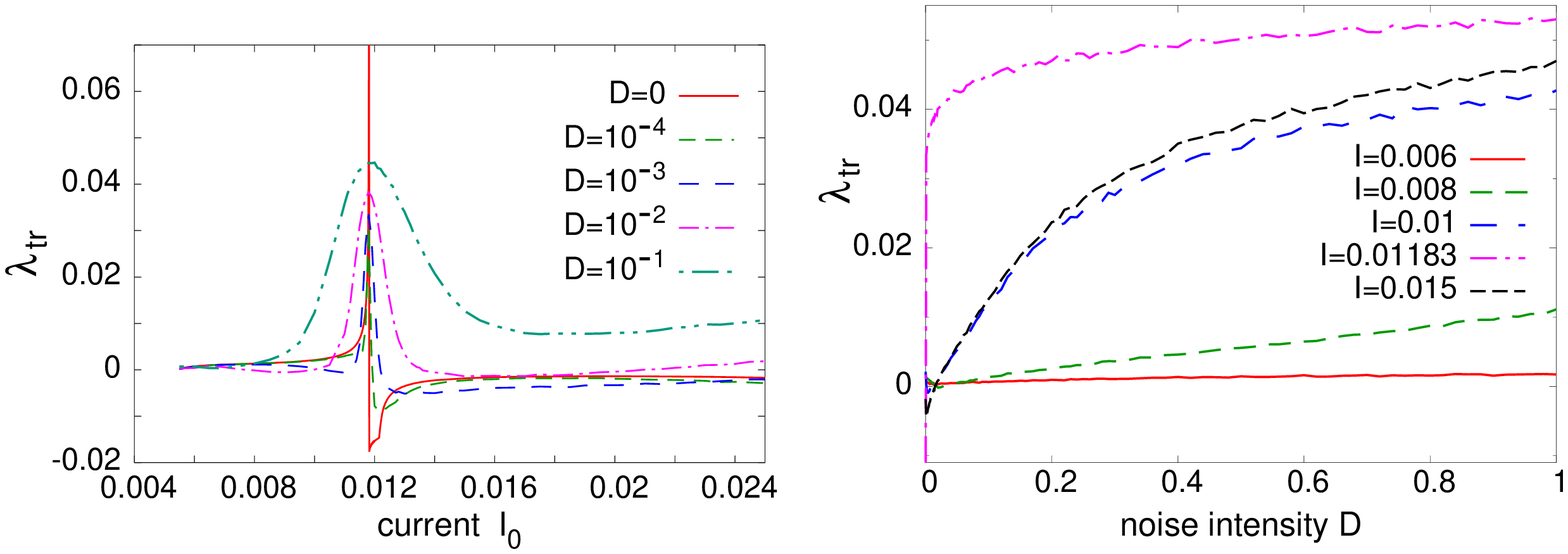}}
\caption{Circuit with purely resistive load under noisy currrent.
Transversal Lyapunov exponent as a function of average current $I_0$
at fixed noise intensity $D$ (left panel) and as a function
of $D$ at fixed $I_0$ (right panel).
Parameter values: see Fig.~\ref{fig_periods}a). }
\label{fig_resistive_along_c}
\end{figure}

We start with the purely resistive circuit.
Dependence of the transversal  Lyapunov exponent on $I_0$ and
$D$ is shown in Fig. \ref{fig_resistive_along_c}.

Recall that in the deterministic case $D$=0,
the evaporation multiplier $\mu_{\mbox{e}}$ of the periodic solution
diverges at the value
of $I$ corresponding to homoclinicity. As a consequence, 
the transverse Lyapunov exponent $\lambda_{\mbox{tr}}$, 
that for a closed orbit with period $T$ equals  
$\log(|\mu_{\mbox{e}}|)/T$, tends to the largest eigenvalue
of the Jacobian matrix at the equilibrium.
In presence of noise, the system spends less time in the vicinity 
of the equilibrium where the instability rates are especially high; 
this results in broadening and softening of
the peak in the dependence of $\lambda_{\mbox{tr}}$ on $I_0$. 
This tendency is apparent in the left panel of Fig.~\ref{fig_resistive_along_c}:
the higher the noise intensity $D$, the broader the maximum 
of $\lambda_{\mbox{tr}}$.
Over large intervals of $I_0$ the weak noise shifts 
$\lambda_{\mbox{tr}}$ downwards; this results in ranges of $I_0$ 
with mildly negative transversal Lyapunov exponent. The effect,
however, remains local in the phase space and does not seem 
to influence global dynamics of the STO ensembles: 
as already mentioned,
unless the initial distribution of oscillators in the ensemble
is extremely narrow, simulations show no convergence 
to the synchronous 1-cluster solution.
The stronger noise, exemplified in Fig.~\ref{fig_resistive_along_c}
by the curve for $D=10^{-1}$, raises the value of $\lambda_{\mbox{tr}}$
everywhere outside the immediate vicinity of the homoclinic singularity,
and amplifies the instability of the synchronous state.
In the right panel the same transversal Lyapunov exponent
is plotted as a function of 
noise intensity at several fixed values of the average current $I_0$;
except the narrow region adjoining the deterministic case,  
$\lambda_{\mbox{tr}}$ appears to be a roughly monotonically growing 
function of $D$.

\begin{figure}[h]
\centerline{\includegraphics[width=0.3\textwidth]{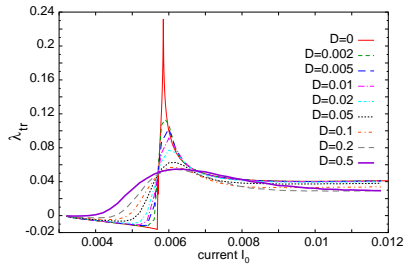}}
\caption{Circuit with the RC load.
Dependence of transversal Lyapunov exponent on the average current $I_0$
at different intensities $D$ of common noise.
Parameter values: see Fig.~\ref{fig_periods}b)}
\label{fig_rc_along}
\end{figure}

Proceeding to the case of the STO stack with the RC load 
(Fig.~\ref{fig_rc_along}), we observe that
here, as well, introduction of the noisy modulation of the common current
broadens and softens the peak of the transversal Lyapunov exponent,
rendering  $\lambda_{\mbox{tr}}$, over the large parameter ranges, positive.
Summarizing, in neither of these two circuit configurations does the common
noise facilitate synchrony.

\subsubsection{STO in the  LC circuit: noise-induced oscillation death}
The situation in presence of the LC load is qualitatively different
from the discussed cases.
\begin{figure}[h]
\centerline{\includegraphics[width=0.5\textwidth]{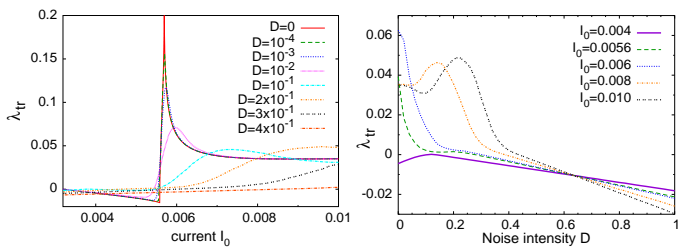}}
\caption{STO in a circuit with LC load and common noise.
Transversal Lyapunov exponent as a function of current $I_0$
at fixed noise intensity $D$ (left panel) and as a function
of noise intensity $D$ at fixed current $I_0$ (right panel).
Parameter values: see Fig.~\ref{fig_periods}c)}
\label{fig_LC_along}
\end{figure}
Fig.~\ref{fig_LC_along} shows the values of the transversal
Lyapunov exponent computed for different combinations
of the average current $I_0$ and noise intensity $D$.
As seen in the left panel,
like in the previously discussed cases, the action of noise
softens and widens the maximum of the transversal
Lyapunov exponent. There is, however, an apparent difference:
the maximum not only becomes broader, but, in contrast
to the cases of resistive load or the circuit with a capacitor, 
it gets shifted from the place of homoclinic bifurcation 
in the deterministic system towards higher values of $I_0$. 
At sufficiently large noise intensity
$D$, the transversal exponent $\lambda_{\mbox{tr}}$ stays negative
in the broad range of $I_0$.
The right panel of Fig.~\ref{fig_LC_along} presents the
same results from the different perspective; variation of 
noise intensity at several fixed values of $I_0$. 
In the right half of the plotted range of $D$, the dependence of
$\lambda_{\mbox{tr}}$ on $D$ becomes virtually linear,
with negative slope. Remarkably, within numerical accuracy all curves 
$\lambda_{\mbox{tr}}(D)$ intersect in the same point.  

Recall that for moderate intensity of noise (in the right panel of 
Fig.~\ref{fig_LC_along} this corresponds to $D>0.4$) 
 simulations of ensemble of noise-driven STOs in the
LC circuit have demonstrated trapping of all oscillators
by the equilibrium state. 
In the context in which all trajectories eventually end up at
the equilibrium, the Lyapunov exponents turn into the
real parts of the eigenvalues of the Jacobian at that equilibrium.
The real part of two leading eigenvalues 
equals
\begin{eqnarray}
\label{instantaneous}
\lambda&=&
-\alpha\left(H_a+H_k+\frac{H_{dz}}{2}\right)+
   \beta\, \overline {(1-w(t))I(t)} \nonumber\\
   &=&
-\alpha\left(H_a+H_k+\frac{H_{dz}}{2}\right)+
   \beta\,I_0\, \overline{\xi(t) w(t)}\sqrt{2D}\nonumber
\end{eqnarray}
where the stochastic variable $w(t)$ (the rescaled time derivative of the voltage $u$), 
is governed by the last of Eq. (\ref{LC_eq_stoch}),
and the overline denotes averaging over time.

To estimate $\overline{\xi(t) w(t)}$ we utilize the fact  
that the equilibrium value of $u$ equals $1-\varepsilon$
and assume that near the equilibrium the variable $w(t)$ obeys the Gaussian distribution
centered at zero. In this way we derive
$$\overline{\xi(t) w(t)}= \frac{\Omega^2(1-\varepsilon)}{2\omega}\sqrt{2D}$$
and, finally, arrive at
\begin{eqnarray}
\label{lyap_noise}
\lambda&=&\beta (I_0-I_H) - \frac{\Omega^2(1-\varepsilon)\beta\,I_0}{\omega}\,D
\nonumber\\[-2ex]
&&\\[-.2ex]
&=& \beta\left(I_0-\frac{\alpha}{\beta}(H_k+H_a+\frac{H_{dz}}{2})\right) 
- \frac{\Omega^2(1-\varepsilon)\beta\,I_0}{\omega}\,D.\nonumber
\end{eqnarray} 
Notably, this expression involves, without exception, all parameters
of the stochastic differential equations (\ref{LC_eq_stoch}).
The last term (recall that $\varepsilon<1$ by definition) 
shows that the noise always lowers the value of $\lambda$.

For $I_0>I_H$, the Lyapunov exponent of the equilibrium 
in the absence of noise is positive. Stochastic trajectories only seldom 
(if at all) visit the neighborhood of the equilibrium, therefore the value 
of this exponent stays local, 
dynamically irrelevant and is not related to the actually observed
value of the leading Lyapunov exponent for generic non-stationary trajectories. 
Increase of the noise intensity weakens the instability, leading
to the gradual decline of $\lambda$. Finally, on crossing the
critical value  
\begin{equation}
\label{value_trap}
D_{\rm trap}=
\,\frac{\omega}{\Omega^2(1-\varepsilon)}
   \left(1-\frac{I_H}{I_0}\right) 
\end{equation} 
the equilibrium gets stabilized and turns into the
global attractor. Henceforth, linearization in its vicinity dominates 
also the global Lyapunov exponent and becomes well visible at its plots. Remarkably,
the value of $D_{\rm trap}$ is a monotonically growing bounded 
function of the average current $I_0$: noise with intensity 
$\displaystyle D>D_g=\frac{\omega}{\Omega^2(1-\varepsilon)}$ 
guarantees trapping of all
oscillators at arbitrary values of the current 
(under employed values of the 
circuit parameters, $D_g\approx 0.635$).

The estimate (\ref{lyap_noise}) turns out to be rather accurate:
in all our simulations for $D>D_{\rm trap}$, the relative discrepancy between the
theoretical prediction (\ref{lyap_noise}) and numerically computed
value of the Lyapunov exponent of the stochastic trajectory never exceeded 0.8\%.

Linear dependence on the noise intensity $D$ explains both the linear character
of the curves in the right part of Fig.~\ref{fig_LC_along} and the fact
of intersection of all curves in the right panel of Fig.~\ref{fig_LC_along}
in the same point\footnote{All curves $\lambda(D)=c+I_0(a-b D)$ with constant
$a,b,c$ and arbitrary $I_0$ intersect at $D=a/b$.}.

\begin{figure}[h]
\centerline{\includegraphics[width=0.35\textwidth]{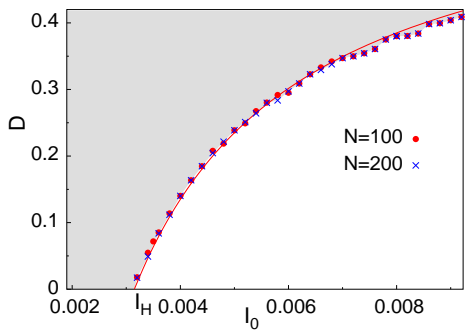}}
\caption{Stabilization border for $N$=100 and $N$=200.
Solid line: theoretical prediction (\ref{value_trap}). Crosses and circles:
numerically determined threshold values of $D$ for the ensembles of, 
respectively, 100 and 200 STOs. Gray background: region of stability for the
equilibrium with $m_x$=1}.
\label{fig_border}
\end{figure}

On the parameter plane spanned by the average current $I_0$ and
the noise intensity $D$ the region of trapping, adjoining the region in which
the equilibrium is stable, lies to the right from $I_{\rm hopf}$  (Fig.~\ref{fig_border});
its lower boundary branches from zero and grows in the direction of $D_g$ at
larger values of $I_0$.

Notably,  this stabilization of the equilibrium by common noise with subsequent
trapping of the ensemble can hardly be called a collective phenomenon:
the ensemble size $N$ enters neither Eq.~(\ref{lyap_noise}) for
the Lyapunov exponent nor the expression (\ref{value_trap}) for
the critical noise intensity. 
A single spin-torque rotator in the LCR circuit with $I_0>I_{\rm H}$ would be
attracted to the equilibrium $m_x$=1 as well, provided that the noise intensity $D$ 
exceeds the threshold (\ref{value_trap}).
Numerically, we estimated the threshold for ensembles of different sizes, 
by maximizing over hundreds of realizations the values of $D$ 
at which macroscopic parts of the ensemble were still not trapped after $t=10^5$.
These critical values
display practically no variation when the ensemble size is doubled
(cf. crosses and circles in Fig.~\ref{fig_border}. 
Only the relaxation time, required for trapping of the \textit{last} ensemble 
element shows a slight increase for the larger $N$.
A single spin-torque rotator in the LC circuit with $I_0>I_{\rm H}$ would be
attracted to the equilibrium $m_x$=1 whenever the noise intensity $D$ exceeds
the threshold (\ref{value_trap}).

\subsection{Discussion}

A natural question is: why does the LC circuit under common noise enable complete 
trapping of the ensemble whereas the other configurations of the circuit fail 
to feature durable stabilization of the equilibrium? The explanation is given by
the way in which the individual and the collective (if present) variables  are affected 
by the common noise.
In the circuit with purely resistive load, governing equations for the angular variables
explicitly include the random term $\xi(t)$. 
Although the instantaneous value of the leading equilibrium eigenvalue
contains terms proportional to $\xi(t)$, 
in the long run these terms average out and bear no influence on the  
overall value $\lambda$ of the Lyapunov exponent
(although, as we have seen, within finite time windows this value can stay negative,
featuring a kind of intermittent stabilization). 
In contrast, equations for the angular variables in the stochastic circuit 
with the capacitor do not contain random terms; there, noise is explicitly
present only in the governing equation for the global (collective) variable $u$ 
(rescaled voltage). This presence, as well,  adds the term $\sim \xi(t)$ 
to the expression for the instantaneous eigenvalue, 
but this term again averages out and does not influence $\lambda$.  
Finally, dynamical description of the circuit with the LC load 
involves noisy terms \textit{both} in the angular variables 
and in one of the global variables. 
As a result, the expression for the instantaneous leading eigenvalue
of the Jacobian (cf. Eq.(\ref{instantaneous})) contains the product of two random terms, 
and the non-zero average value of this product serves for the systematic 
noise-induced shift of $\lambda$.

\section{Conclusion}
The main goal of this paper was to focus on peculiarities of the STOs that strongly affect 
their synchronization properties.  We show that the gluing bifurcation implies divergence
of the transversal Floquet multiplier, responsible for stability of the synchronized cluster.
This phenomenon is intrinsic to the STO dynamics and cannot be mitigated by variations
of the load; the only way to avoid the instability appears to select external control parameters to
be far away of the homoclinic transition. We also analyzed another important factor 
influencing the dynamics of STOs, namely fluctuations of the current through the array.
Here we observed a novel feature of suppression of oscillations, which can be interpreted as
\textit{noise-induced oscillation death}. Here a distribution of the external noisy input between the stack of STOs
and the parallel load is important: only an LC load leads to an effective shift of the Hopf bifurcation
end eventual stabilization of the steady state by noise. 

\noindent\textbf{Acknowledgments}\\
Numerical part of this work conducted by A.P. was
supported by the Russian Science Foundation
(Project No. 17-12-01534). M.Z. was supported by DFG (grant PI 220/17-1).

\end{document}